\documentclass[10pt]{iopart}

\usepackage{graphicx}

\usepackage{hyperref} % For hyperlinks in the PDF

\newcommand{\pr}{\partial}
\newcommand{\rta}{\rightarrow}

\newcommand{\ep}{\epsilon}

\newcommand{\beq}{\begin{equation}}
\newcommand{\eeq}{\end{equation}}

\newcommand{\ben}{\begin{enumerate}}
\newcommand{\een}{\end{enumerate}}

\begin{document}

\title[Localized-to-itinerant transitions]{A toy model for localized to itinerant electron transitions}

\author{Navinder Singh}

\address{Theoretical Physics Division, Physical Research Laboratory, Ahmedabad. India. PIN: 380009.}
\ead{navinder.phy@gmail.com; navinder@prl.res.in}
\vspace{10pt}
\begin{indented}
\item[]3rd May 2021
\end{indented}

\begin{abstract}
A toy statistical model which mimics localized-to-itinerant electron transitions is introduced. We consider a system of two types of  charge carriers: (1) localized ones, and (2) itinerant ones. There is chemical equilibrium between these two.  The solution of the model shows that there is a crossover, at a specific value of a local interaction energy parameter $J$, at which carriers transfer from being itinerant to a localized/mixed valence state. The model has a very crude analogy to the observed  $\gamma\rta\alpha$ transition in metal cerium.
\end{abstract}

\vspace{2pc}
\noindent{\it Keywords}: A toy model for localized to itinerant electron transitions; transition metals and their oxides; ideal gas in contact with an adsorbing surface; mixed valence systems.
%\maketitle


\section{Definition of the problem}
%%%%%%%%%%%%%%%%%%%%%%
There is a very interesting problem in Kubo's book on statistical mechanics\cite{kubo}. The problem is related to chemical equilibrium between an ideal molecular gas and localized molecules adsorbed on an adsorbent surface (figure 1). Depending upon the temperature and pressure of the gas and on the properties of the adsorbent surface, a fraction of molecules get adsorbed on the surface. The problem is to determine the ratio of adsorbed molecules to the number of adsorbing sites (the covering ratio) as a function of the physical parameters of the gas (like temperature). 

The solution of the problem turns out to be
\begin{figure}[!h]
\begin{center}
\includegraphics[height=6cm]{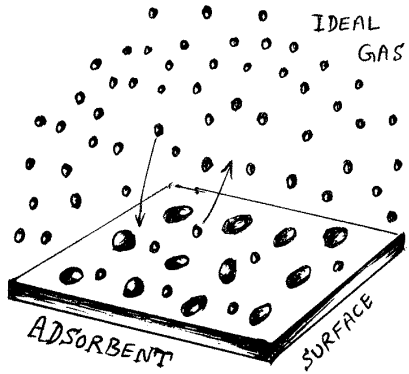}
\caption{An ideal gas in chemical equilibrium with some of its molecules adsorbed on an adsorbent surface.}
\end{center}
\label{f1}
\end{figure}

\beq
Covering~Ratio= \frac{p}{p+p(T)},~~~~~~\\ p(T) = k_B T e^{-\frac{\ep_0}{k_BT}} \left(\frac{2 \pi m k_B T}{h^2}\right)^{3/2}
\eeq
Here $p$ is the pressure of the gas, and $-\ep_0$ is the energy lowering when a molecule adsorbs on the surface. The covering ratio first increases with pressure and then saturates to one at sufficiently high pressure determined by the temperature through $p(T)$ (refer to equation (1)).

We generalize this problem. Instead of a classical gas we take a quantum gas of electrons where the chemical potential is not determined through pressure but by using the number of electrons in the gas (itinerant) phase. We consider a system in which we can divide the charge carriers into two sub-categories: (1) localized ones, and (2) itinerant ones. There is chemical equilibrium between these two. The problem which we would like to solve now is this: Is there a critical value of the ``local adsorbing energy" $J$ below which some of the itinerant carriers goes into localized ones, and how this energy scale $J$ depends upon the temperature? We will notice that the problem and the properties of its solution mimic the itinerant to localized/mixed valence transitions found in some transition metals and their compounds. However, we stress that this is just a "toy model" and this exercise is done from purely academic interest.

To make the ideas concrete assume that we have an $f-$electron system. Electrons can be localized in $f-$orbitals (sequestered deep inside their atoms) or they can become itinerant by migrating to a fat $s-$band (say) via some coupling energy scale. There is a chemical equilibrium between these two types of carriers. We want to learn how do electrons migrate as temperature or other parameters of the model are changed.

\section{Solution of the problem}
Let there be $N$ sites per unit area on a two dimensional lattice. Assume that there is one unpaired electron at each site (or orbital) to start with. The energy cost to localize a given electron at a site is $J$.  At a given temperature $T$ and at a given value of $J$ there will be both itinerant electrons (that form a Fermi surface) and localized electrons on the lattice. We consider that there is no energy gap between the bottom of the Fermi distribution of mobile electrons and the local energy scale $J$.  We divide this system into two components: (1) localized electrons, and (2) itinerant ones, both in mutual thermodynamical equilibrium with each other. Let $n$ out of $N$ electrons form temporary localized states. Then their energy is given as $J n$ (energy cost for one electron to localize in an orbital is $J$). The canonical the partition function is:
\beq
Z_n = \frac{(N)!}{n! (N-n)!} e^{-\beta J n}.
\eeq 
If we denote the chemical potential as $\mu$, then the grand partition function is given by

\beq
Q = \sum_{n=0}^N Z_n e^{\beta \mu n} = \sum_{n=0}^N \frac{N!}{n! (N-n)!} e^{\beta(\mu- J) n} 1^{N-n} = (1+e^{\beta(\mu-J)})^N.
\eeq

The average number of the localized electrons ($N_{loc}$) is:

\beq
N_{loc} = \frac{1}{\beta}\frac{\pr}{\pr \mu}Log(Q) = N \frac{e^{\beta(\mu-J)}}{e^{\beta(\mu-J)}+1}.
\eeq

And the number of electrons in the mobile (itinerant state) is $N_{iti} = N-N_{loc}$:

\beq
N_{iti} = N \frac{1}{e^{\beta(\mu-J)}+1}.
\eeq

Now, these itinerant electrons form a Fermi sphere. Therefore, we must have in 2D:

\beq
N_{iti}(T) = \frac{2}{A}\sum_k f_k = 2 \int d^2k \frac{1}{e^{\beta(\ep_k -\mu)}+1}.
\eeq

Here, $f_k$ is the Fermi function, and $\ep_k$ is the energy of a free electron in $k$ state. $A$ is the area of the lattice. Plugging the value of $N_{iti}$ from equation (5) in the above equation, and after some simple algebra, we get the main result:

\beq
\frac{1}{e^{\beta(\mu-J)}+1} = \frac{\alpha}{\beta}Log[1 +e^{\beta\mu}].
\eeq

\begin{figure}[!h]
\begin{center}
\includegraphics[height=4cm]{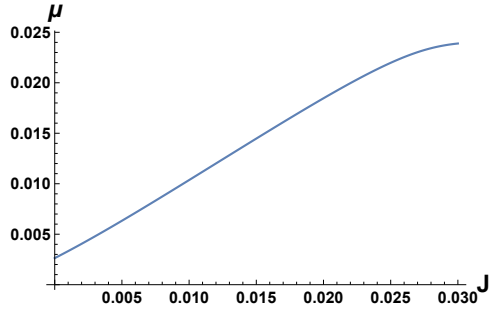}
\caption{Chemical potential as a function of $J$, at $T=15~K$, and $N= 4\times10^{18}/meter^2$.}
\end{center}
\label{f2}
\end{figure}

Here, $\alpha = \frac{4 \pi m}{\hbar^2 N}$. This equation needs to be numerically solved to find $\mu$ for given values of $T$, $J$, and $N$. Then, from $\mu(T)$ we can find out the temperature dependence of $N_{loc}$ and $N_{iti}$. However, at present,  we are interested in the $J$ dependence of $\mu, ~N_{iti}$, and $N_{loc}$. The problem of temperature dependence is treated in the next section. Numerical computation of $\mu(J)$ is presented in figure (\ref{f2}) for a fixed value of temperature and $N$. It is positive, as it should for fermions, and increases with increasing $J$. This is also understandable: higher value of $J$ leads to higher energy cost to localize an electron, thus they tend to be itinerant and form a bigger Fermi surface at a larger value of $\mu$.

\begin{figure}[h!]
\begin{center}
\begin{tabular}{cc}
\includegraphics[width =5cm]{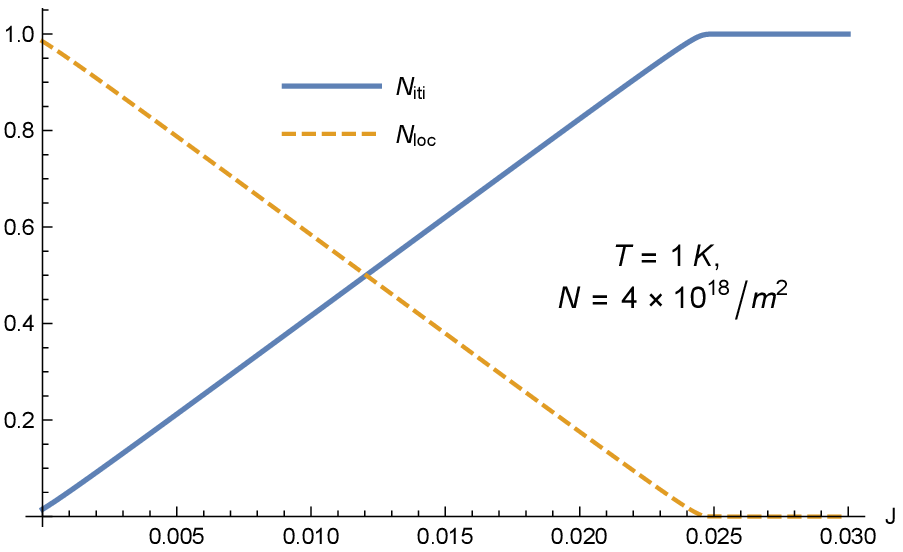}&
\includegraphics[width =5cm]{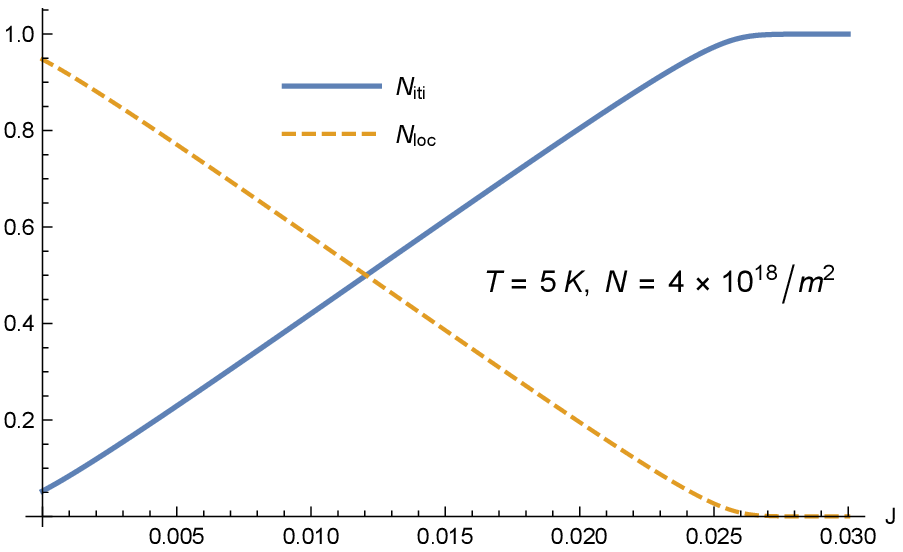}\\
(a)&(b)\\
\end{tabular}
\end{center}
\caption{(a) $J$ dependence of $N_{iti}/N$ and $N_{loc}/N$ at $T= 1~K$. (b).  $J$ dependence of $N_{iti}/N$ and $N_{loc}/N$ at $T= 5~K$.  $N= 4\times10^{18}/meter^2$.}
\label{f3}
\end{figure}

From the knowledge of $\mu(J),~N_{iti}$ and $N_{loc}$ can be computed from equation (5) [$N_{iti} = N-N_{loc}$]. Figures \ref{f3} (a) and (b) represents $J$ dependence of $N_{iti}(J)$ and $N_{loc}(J)$ at temperature $1~K$ and $5~K$. It is clear that for $J>0.025~eV$ all the electrons are in the itinerant phase (and form a Fermi distribution). All the local sites are left vacant. But, as the value of $J$ is reduced from $0.025~eV$ some electrons migrate from the fermi distribution and stay on the local sites. Below $J = 0.025~eV$ we have a mixed valence phase (some local sites are vacant and some having an electron).  For the $J$ value of $0.012 ~eV$ we have $50-50$ mixture. And very low values of J, we majority of the electrons localized.

This apparently surprising behaviour can be easily understood. At large $J$ local repulsion is large, and it is energetically favorable for the electrons to form a Fermi distribution. But for low value of $J$ local repulsion is less, and some of the electrons migrate from the top of the Fermi distribution (thereby reducing their K. E.) to the local sites. There is noting sacred for the value $0.025~eV$ of $J$. This number is governed by the total number of electrons (number of sites) taken at the start (For numerical computation we imagine a 2D lattice with lattice constant $5\AA$, and each site providing one electron, so that $N = 4\times10^{18}~meter^{-2}$). More number of free electrons (and sites) means a bigger Fermi surface and more K. E. Then this crossover happens even at a larger value of $J$.

\begin{figure}[h!]
\begin{center}
\begin{tabular}{cc}
\includegraphics[width =6cm]{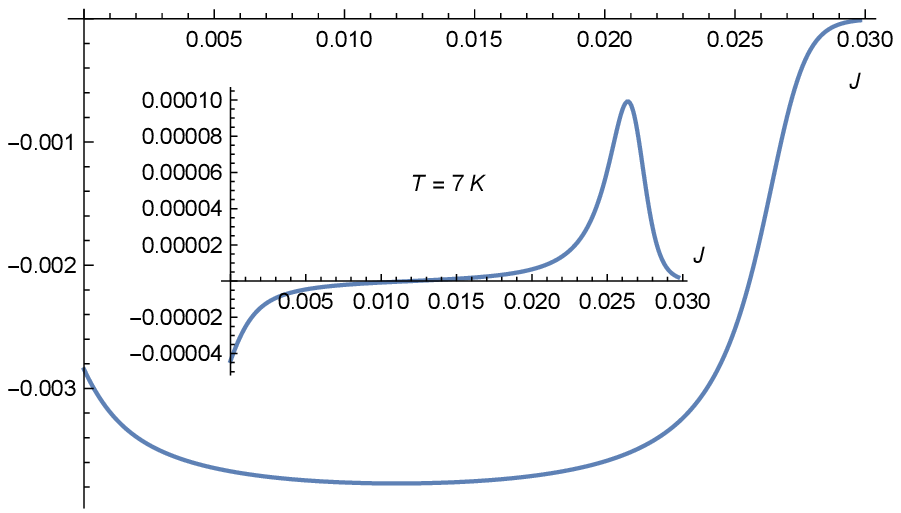}&
\includegraphics[width =6cm]{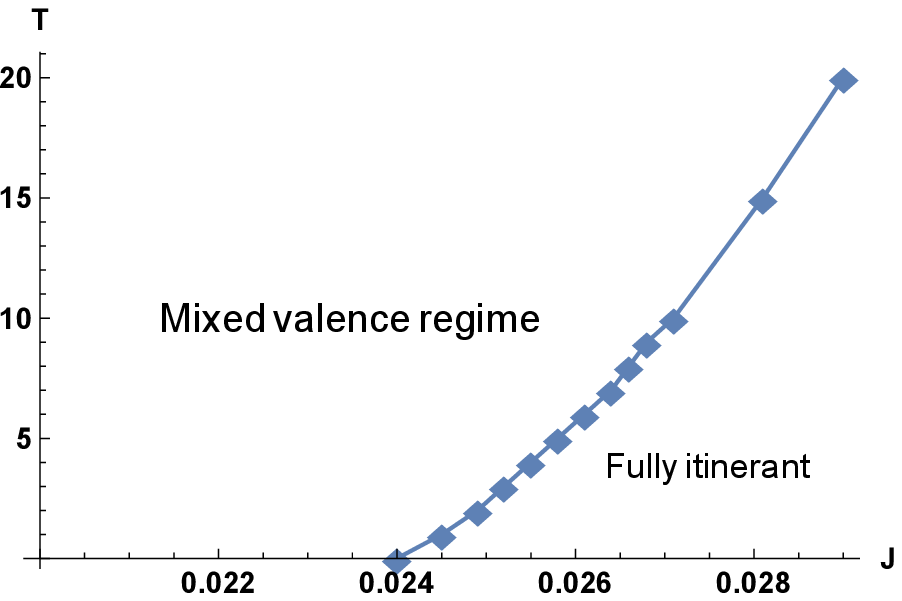}\\
(a)&(b)\\
\end{tabular}
\end{center}
\caption{(a) Slope of $N_{loc}$ as a function of $J$. Inset shows the double derivative of $N_{loc}$.  (b) The resulting phase diagram   $N= 4\times10^{18}/meter^2$.}
\label{f4}
\end{figure}

Next we study how this critical value of $J$ that is $J_c$ (above which we have itinerant electrons, and below which we have a mixed valence system) evolve with temperature. To identify $J_{c}$ we look at the derivatives of $N_{iti}$ or $N_{loc}$ wrt $J$. We define $J_c$ at which the slope of $N_{iti}$ or $N_{loc}$ exhibits a maximum change (figure \ref{f4}  (a) inset shows the double derivative of $N_{loc}$). This is the way we determine $J_c$. The value of $J_c$ is temperature dependent. It is determined at various temperatures using the above method, and the resulting phase diagram is plotted in figure \ref{f4} (b). It is the main result of this work.

From the phase diagram (figure \ref{f4} (b)) it is clear that towards the right hand side of $J_c-T$ boundary we have fully itinerant system, and towards the left of it we have a mixed valence system (some electrons are itinerant and some are localized). This the the solution of the problem which we defined in section (1).

Now the question is whether this toy model has any resemblance with a real localized-to-itinerant electron transitions observed in transition metals and their oxides or in heavy fermion materials? As far as crude analogy is concerned the answer is yes! The $\gamma\rta\alpha$ phase transition in cerium metal has such similarities\cite{1,2,3,4,5}. If we look at the phase diagram in figure 1 in reference\cite{1} there is a phase boundary between the $\alpha-$phase (itinerant electron phase) and $\gamma-$phase (localized electron phase) as a function of pressure (the tuning parameter). The reason might be that as the pressure is increased it becomes more energetically costly for the $f-$electrons to stay in sequestered $f-$orbitals and they tend to migrate out of these, and move into the itinerant band. Thus the present phase diagram (figure 4(b)) mimics the one in figure 1 of reference\cite{1}. We stress that this toy model has only crude resemblance but it will initiate more studies in this field.   One has to investigate quantitatively how pressure change effects $J$ (the local repulsion) and what are the exact roles of Kondo coupling and Mott physics in the system.

\section{Approximate analytical solution for $\mu$ in a special case, and other extensions}
An explicit equation for $\mu$ can be obtained in a special case of semiclassical approximation. Suppose that the number of available $k$ states is much greater than the total number of electrons. Then the average number of electrons per quantum state will be much less than one, and in this case Fermi distribution reduces to Boltzmann distribution. One can write equation (6) as 

\beq
N_{iti} = \frac{4\pi m}{\hbar^2} \int_0^\infty d\ep e^{-\beta(\ep -\mu)}.
\eeq
Solving for $\mu$, using equation (5), one can easily show

\beq
\mu = k_B T \log\left[\frac{1}{2}e^{\beta J} \left(\sqrt{1+\frac{N\hbar^2}{\pi m k_B T}e^{-\beta J}}-1\right)\right].
\eeq
From which an explicit expressions for $N_{iti}(T)$ and $N_{loc}(T)$ can be easily written down.

\begin{figure}[h!]
\begin{center}
\begin{tabular}{cc}
\includegraphics[width =6cm]{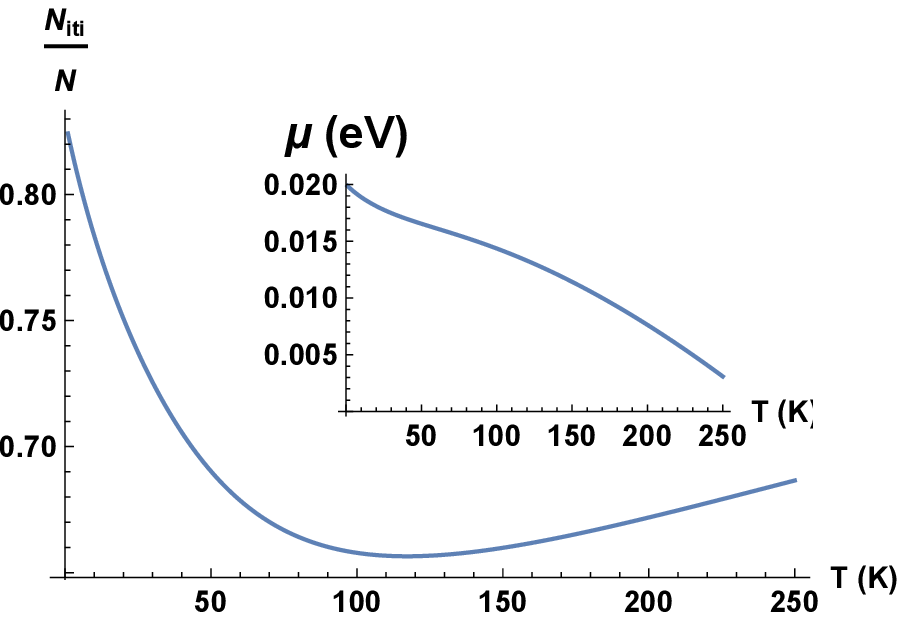}&
\includegraphics[width =6cm]{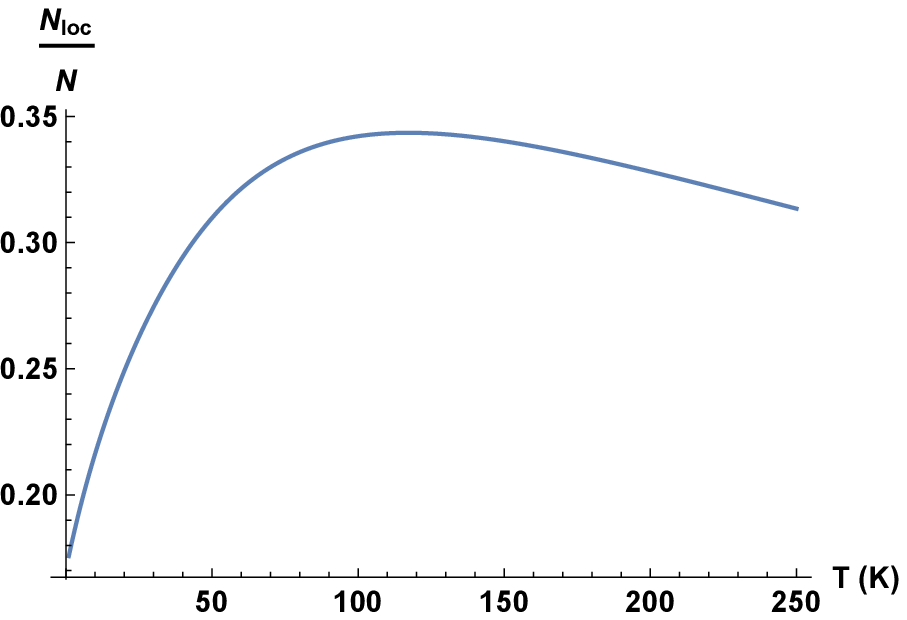}\\
(a)&(b)\\
\end{tabular}
\end{center}
\caption{(a) $\frac{N_{iti}}{N}$ as a function of temperature (inset shows the temperature dependence of the chemical potential).  (b) $\frac{N_{loc}}{N}$ as a function of temperature. It shows maximum around $T^* \simeq100 ~K$ ( $T^*$ can be obtained from the numerical solution of $\frac{d\mu(T^*)}{dT^*} = \frac{\mu(T^*)-J}{T^*}$). Here $N= 4\times10^{18}/meter^2$.}
\label{f6}
\end{figure}

Next, we study the temperature dependence of $N_{iti}$ and $N_{loc}$. A wide variety of monotonic behaviour of $N_{iti}$ and $N_{loc}$ as a function of temperature is observed at various values of $J$ and $N$.  A specifically interesting behaviour occurs when $J\simeq \frac{1}{\alpha}$\footnote{$N_{iti}(T)$ will have maximum or minimum when $\frac{d N_{iti}(T)}{dT} = 0$. This condition leads to a differential equation: $\frac{d\mu(T^*)}{dT^*} = \frac{\mu(T^*)-J}{T^*}$ solution of which gives $T^*$ the temperature at which extremum occurs.  Also the solution of this differential equation at $T=0$ leads to $\mu = J$. If we now simplify the equation (7)--the defining equation of $\mu$--in the zero temperature limit we get $\mu\simeq \frac{1}{\alpha}$ ($\mu$ is less than $J$ as $\mu = J - C k_B T$). Therefore, whenever $J \simeq \frac{1}{\alpha}$ one obtains extremum in the temperature evolution of the populations.}. In such cases the populations $N_{iti}(T)$ and $N_{loc}(T)$ pass through minimum and maximum respectively! That is the number of localized electrons becomes maximum at a particular temperature, it decreases when the temperature is either reduced or increased! This is quite interesting and depicted in figures (\ref{f6}).

The reason for this kind of behaviour is that there is a competition between K. E. of free electrons forming the Fermi sphere and P.E. of electrons those are localized ($J\times$ number of localized electrons). When there is a balance between these two, populations, $N_{loc}$ etc, pass through extrema. 

\section{Conclusion}

Even when the premises on which the model is build are simple, but properties of its solution are very interesting and mimic the localized to itinerant electron transitions observed in transition metals. The model can be easily extended to 1-D case (Luttinger liquid case) and the more common 3-D case. It can serve as a prototype, and can initiate more serious studies especially in the $\gamma\rta\alpha$ phase transition in cerium metal which has such similarities. 

%\section{Data Availability}
%No data is generated or used in this study.

%\section{Author Contribution}
%This is a single author paper, and all the work is done by the presenting author.

%\section*{Acknowledgments}

\section*{References}

%----------------------------------------------------------------------------------------
%	REFERENCE LIST
%----------------------------------------------------------------------------------------

%----------------------------------------------------------------------------

\end{document}